\documentclass[11pt,fleqn]{article} 
\usepackage{graphicx}
\usepackage{textcomp} 
\usepackage{epstopdf}
\usepackage{latexsym}
\usepackage{amsmath}
\usepackage{amssymb}
\usepackage{bm}
\usepackage[mathcal]{euscript}
\usepackage{slashed}
\usepackage[utf8]{inputenc}

\usepackage[top=35mm, bottom=30mm, left=25mm, right=25mm]{geometry}

\numberwithin{equation}{section}

\usepackage{marvosym}

\usepackage{indentfirst}

\newcommand\blfootnote[1]{
  \begingroup
  \renewcommand\thefootnote{}\footnote{#1}
  \addtocounter{footnote}{-1}
  \endgroup
}

\usepackage{hyperref}
\hypersetup{
        unicode,	
        colorlinks,
        citecolor=blue,
        linkcolor=blue, 
        urlcolor=red,
        bookmarksopen=true,
        bookmarksopenlevel=\maxdimen,
      }

\def\gl#1#2{\ifmmode \mathrm{GL}(#1; {\bf #2}) \else $\mathrm{GL}(#1; {\bf #2})$\fi}
\def\sl#1#2{\ifmmode \mathrm{SL}(#1; {\bf #2}) \else $\mathrm{SL}(#1; {\bf #2})$\fi}
\def\so#1{\ifmmode \mathrm{SO}({#1}) \else $\mathrm{SO}(#1)$\fi}

\def\sp#1#2{\ifmmode \mathrm{Sp}(#1; {\bf #2}) \else $\mathrm{Sp}(#1; {\bf #2})$\fi}
\def\usp#1{\ifmmode \mathrm{USp}(#1) \else $\mathrm{USp}(#1)$\fi}
\def\spin#1{\ifmmode \mathrm{Spin}(#1) \else $\mathrm{Spin}(#1)$\fi}
\def\su#1{\ifmmode \mathrm{SU}({#1}) \else $\mathrm{SU}(#1)$\fi}

\def\double #1{#1{\hbox{\kern-2pt $#1$}}}

\mathcode`\*="702A                  
\def\half{{\textstyle{1\over{\raise.1ex\hbox{$\scriptstyle{2}$}}}}}

\def \p{\partial}
\def \a{\alpha}
\def \b{\beta}

\def \d{\delta}

\def \g{\gamma}
\def \l{\lambda}

\def\r{\rho}
\def \L{\Lambda}
\def \Lb{\overline\Lambda}

\def \o{\omega}

\def \O{\Omega}
\def \t{\theta}

\def \psu{{\mathfrak{psu}}}

\def \sJ{{\mathsf J}}
\def \sN{{\mathsf N}}
\def\ua{\underline{a}}
\def\ub{\underline{b}}
\def\uc{\underline{c}}
\def\ud{\underline{d}}
\def \ue{\underline{e}}
\def\uab{\underline{ab}}
\def\ucd{\underline{cd}}
\def\ubc{\underline{bc}}
\def\uba{\underline{ba}}

\def \N{\nabla}

\begin{document}

\begin{flushright}
\makebox[0pt][b]{}
\end{flushright}

\vspace{40pt}
\begin{center}
{\LARGE A superfield realization of the integrated vertex operator \\ in an $AdS_5\times S^5$ background}

\vspace{30pt}
Osvaldo Chandia${}^\clubsuit$ and Brenno Carlini Vallilo${}^{\spadesuit}$
\vspace{40pt}

{\em ${}^\clubsuit$ Departamento de Ciencias, Facultad de Artes Liberales \& UAI Physics Center,\\ Universidad Adolfo Ib\'a\~nez, Chile
\vspace{10pt}
\\${}^{\spadesuit}$ Departamento de Ciencias F\'{\i}sicas, Universidad Andres Bello, \\
Sazie 2212, Santiago, Chile
}\\

\vspace{60pt}
{\bf Abstract}
\end{center}
The integrated massless vertex operator in an $AdS_5\times S^5$
background in the pure spinor formalism is constructed in terms of
superfields.

\blfootnote{\\
${}^\clubsuit$\href{mailto:ochandiaq@gmail.com}{ochandiaq@gmail.com}\\
${}^{\spadesuit}$ \href{mailto:vallilo@gmail.com}{vallilo@gmail.com} }

\setcounter{page}0
\thispagestyle{empty}

\newpage

\tableofcontents

\parskip = 0.1in

\section{Introduction}

A covariantly quantizable action for the type IIB
superstring was proposed some time ago using the pure spinor
formalism \cite{Berkovits:2000fe,Berkovits:2000yr}. In this formalism, the
physical states are defined as cohomology elements of a nilpotent
BRST-like charge. The light-cone spectrum obtained \cite{Berkovits:2000nn}
coincides with the answer one for a superstring in flat ten
dimensional space. This was done using non-covariant methods.
The covariant description of states and vertex operators is much less
known.  Only the first massive state of the open superstring was
studied covariantly \cite{Berkovits:2002qx}. An attempt to describe a
specific massive state in the $AdS_5\times S^5$ background was made in
\cite{Vallilo:2011fj}.

The condition of having the BRST-like symmetry in generic
background fields constrains them to be on-shell.
This was done for supregravity \cite{Berkovits:2001ue}.
For the type IIB superstring in an $AdS_5\times S^5$ background,
it was shown that this is exact for all orders in perturbation theory
\cite{Berkovits:2004xu}. Fluctuations around a given background are
the vertex operators of the theory. They are also necessary to compute
amplitudes. The problem of finding explicit
descriptions of specific physical states of the superstring in
curved backgrounds is very difficult both in RNS of GS-like
descriptions. Apart from BMN limits of $AdS$ backgrounds
\cite{Metsaev:2002re,Chandia:2014wca},
as far as we know, the only  construction of a
vertex operator for some specific state
in curved backgrounds has been done very recently in
\cite{Adamo:2017sze} for the ambitwistor RNS superstring
\cite{Mason:2013sva,Adamo:2014wea} in a plane wave background.

The structure of the unintegrated vertex operator for $AdS_5\times
S^5$ was first described in \cite{Berkovits:2000yr}. A formal construction
for the integrated vertex operator in this case was done in
\cite{Chandia:2013kja} using the string Lax pair
\cite{Vallilo:2003nx,Mikhailov:2012uh,Chandia:2016ueo}. The explicit
form of the integrated vertex operator was not considered in this reference.
The purpose of this paper is to construct the explicit superfields
of the integrated vertex operator and
their constraints. Due to the lack of a fundamental $b$ ghost,
the relation between the unintegrated vertex operator $U$ and
the integrated vertex operator $V$ are better described by a descent procedure
\begin{align}
QU=0,\quad \p U= QW,\quad \bar\p U = Q\bar W,\quad QV=\p\bar W - \bar\p W ,
\label{chain0}
\end{align}
where $Q$ is the pure spinor BRST-like charge and $(W,\bar W)$ are
operators defined by these equations.

This paper is organized as follows. Section \ref{2} contains a short
review of the pure spinor formalism and a detailed description of its
massless vertex operators. In section \ref{3} we begin with the
description of the pure spinor version of type IIB superstring
in an $AdS_5\times S^5$ background. After it we discuss the
unintegrated vertex operator, pointing out some similarities
with the flat space case. In section \ref{4} we find the chain of
operators (\ref{chain0}) for $AdS_5\times S^5$ background with all
superfields containing the physical fluctuations. We conclude the work
in section \ref{5}, discussing future problems and possible
applications.

\section{Massless vertex operators in flat space
  for type II superstring}\label{2}

We will begin with a short review of the pure spinor type II superstring in flat
space. The closed string vertex operator was studied in detail in
\cite{Grassi:2004ih}, here we will review some aspects which will be relevant later. The fundamental
variables are those of $N=2$ ten dimensional
superspace $(X,\t,\bar\t)$ plus the conjugate momenta of the odd
variables $(p,\bar p)$ and a set of ghosts. The action is
\begin{align}\label{S0}
S_0=\int d^2z \left( \frac12 \p X_m \bar\p X^m + p_\a \bar\p \t^\a + \bar p_{\bar\a} \p \bar\t^{\bar\a} + \o_\a \bar\p \l^\a + \bar\o_{\bar\a} \p \bar\l^{\bar\a}\right) ,
\end{align}
and the BRST symmetry is generated by
\begin{align}\label{Q0}
Q=\oint\left( \l^\a d_\a + \bar\l^{\bar\a} \bar d_{\bar\a}\right) ,
\end{align}
where
\begin{align}
d_\a = p_\a + \frac12 (\g^m\t)_\a (\p X_m - \frac14 (\t\g_m\p\t)),\quad \bar d_{\bar\a} = \bar p_{\bar\a} + \frac12 (\g^m\bar\t)_{\bar\a} (\bar\p X_m- \frac14   (\bar\t\g_m\bar\p\bar\t)) ,
\label{d0}
\end{align}
where the $\g^m$'s are the symmetric $16\times 16$ gamma matrices in ten dimensions. The BRST charge is nilpotent when the ghosts $(\l,\bar\l)$ satisfy
\begin{align}
  \l\gamma^m\l=0=\bar\l\gamma^m\bar\l.
\end{align}
These conditions also imply that the anti-ghosts are defined up to
\begin{align}\label{gaugeOmegas}
  \delta \omega_\a = a_m(\gamma^m\l)_\a,\quad
  \delta \bar\omega_{\bar\a} = b_m(\gamma^m\bar\l)_{\bar\a},
\end{align}
for any local parameters $(a_m,b_m)$.

It is useful to work with supersymmetric combinations of the
world-sheet variables. They are the $d, \bar d$ defined above,
world-sheet derivatives of $\t,\bar\t$ and
\begin{align}
\Pi^m=\p X^m+\frac12(\t\g_m\p\t)+\frac12(\bar\t\g^m\p\bar\t),\quad \bar\Pi^m=\bar\p X^m+\frac12(\t\g_m\bar\p\t)+\frac12(\bar\t\g^m\bar\p\bar\t) .
\end{align}
The BRST transformations of these supersymmetric invariants are given by
\begin{align}
 Q \Pi^m &= (\l\g^m\p\t)+(\bar\l\g^m\p\bar\t),\quad Q d_\a=-(\l\g_m)_\a
  \Pi^m,\cr
   Q\p\t^\a& = \p\l^\a,\quad Q\o_\a=d_\a,\quad Q\l^\a=0 ,\label{Qleft}\\
Q \bar\Pi^m &= (\l\g^m\bar\p\t)+(\bar\l\g^m\bar\p\bar\t),\quad Q \bar
  d_{\bar\a}=-(\bar\l\g_m)_{\bar\a} \bar\Pi^m,\cr
  Q\bar\p\bar\t^{\bar\a} &= \bar\p\bar\l^{\bar\a},\quad Q\bar\o_{\bar\a}=\bar d_{\bar\a},\quad Q\bar\l^{\bar\a}=0 .
\label{Qright}
\end{align}
Note that on-shell $\l$ only appears in the holomorphic sector and
$\bar\l$ only appears in the anti-holomorphic sector. This is due to
the fact that $Q$ is only part of a much larger symmetry generated by
the holomorphic and anti-holomorphic currents $j=\l^\alpha d_\alpha$
and $\bar j=\bar\l^{\bar\alpha} \bar d_{\bar\alpha}$.

The BRST transformation of any superfield is
$Q\Psi(X,\t,\bar\t)=\l^\a D_\a\Psi +\bar\l^{\bar\a} D_{\bar\a} \Psi$,
where $D_\a=\p_\a+\frac12 (\g^m\t)_\a\p_m$ and
$D_{\bar\a}=\p_{\bar\a}+\frac12 (\g^m\bar\t)_{\bar\a}\p_m$. The algebra of theses superspace covariant derivatives is given by
\begin{align}
\{D_\a,D_\b\}=\g^m_{\a\b}\p_m,\quad
  \{D_{\bar\a},D_{\bar\b}\}=\g^m_{\bar\a\bar\b}\p_m,\quad
  \{D_\a,D_{\bar\b}\}=0.
\end{align}

Physical states are defined to be in the cohomology of $Q$.
The massless states are described by the unintegrated vertex operator
with vanishing classical dimension\footnote{ A more general form of
  the vertex operator is given by
\begin{align}
\nonumber
 U=\l^\a \bar\l^{\bar\b} A_{\a\bar\b}(X,\t,\bar\t) +\half
  \l^\alpha\l^\beta A_{\alpha\beta}(X,\t,\bar\t) +\half
  \bar\l^{\bar\alpha}\bar\l^{\bar\beta} A_{\bar\alpha\bar\beta}(X,\t,\bar\t).
\end{align}
Although this for is useful for some applications \cite{ Mikhailov:2009rx,Bedoya:2010qz}, all physical states
can be described by the gauge fixed version (\ref{U0}) }
\begin{align}\label{U0}
  U=\l^\a \bar\l^{\bar\b} A_{\a\bar\b}(X,\t,\bar\t).
\end{align}
Since $U$ is in the cohomology of $Q$, its gauge transformation is
given by
\begin{align}
  \delta U = Q\Lambda = Q(\l^\alpha \Lambda_\alpha
  +\bar\l^{\bar\alpha}\bar\Lambda_{\bar\alpha})=
  \l^\a \bar\l^{\bar\b}( D_{\alpha}\L_{\bar\b}+D_{\bar\beta}\L_\a) +
  \l^\alpha\l^\beta D_\a \L_{\b}+
  \bar\l^{\bar\alpha}\bar\l^{\bar\beta}D_{\bar\alpha}\L_{\bar\beta}.
\end{align}
In order to preserve the original form of $U$ the parameters
$(\Lambda_\alpha,\bar\Lambda_{\bar\alpha})$ are constrained to satisfy
\begin{align}\label{condL}
  D_{(\alpha} \Lambda_{\beta)} = \gamma^m_{\alpha\beta}\Lambda_m,\quad
  D_{(\bar\alpha} \bar\Lambda_{\bar\beta)} =
  \gamma^m_{\bar\alpha\bar\beta}\bar\Lambda_m.
\end{align}
These conditions resembles the equations that define two on-shell
vector multiplet in ten dimensions. The main difference is that
here $(\L_\a,\bar\L_{\bar\a})$ are functions of $N=2$ superspace
variables. From them we can also obtain
\begin{align}
  D_\a\L_m-\p_m\L_\a=(\g_m)_{\a\b}\L^\b,\quad
  D_\a\L^\b=\frac14(\g^{mn})_\a{}^\b \p_{[m}\L_{n]},\label{chainL}\\
  D_{\bar\a}\bar\L_m-\p_m\bar\L_{\bar\a}=(\g_m)_{\bar\a\bar\b}\bar\L^{\bar\b},\quad
  D_{\bar\a}\bar\L^{\bar\b}=\frac14(\g^{mn})_{\bar\a}{}^{\bar\b} \p_{[m}\bar\L_{n]}.
\label{chainLbar}
\end{align}
The first components of $(\L_m,\bar\L_m)$ are related to the
diffeomorphism parameters and gauge transformation of the Kalb-Ramond
field and $(\L^\alpha,\bar\L^{\bar\alpha})$ are the local
supersymmetry parameters. We will see this explicitly later.

The condition that $U$ is BRST closed implies
\begin{align}
  QU &=(\l^\a D_\a +\bar\l^{\bar\a} D_{\bar\a})\l^\b \bar\l^{\bar\b}
  A_{\b\bar\b}(X,\t,\bar\t)\cr &=
  \l^\b \bar\l^{\bar\b}\l^\a D_\a A_{\b\bar\b}(X,\t,\bar\t) +
  \l^\b \bar\l^{\bar\b}\bar\l^{\bar\a} D_{\bar\a} A_{\b\bar\b}(X,\t,\bar\t)=0
\end{align}
which is solved by
\begin{align}
  D_{(\a} A_{\b)\bar\g}=\g^m_{\a\b} A_{m\bar\g},\label{DA}\\
  D_{(\bar\a} A_{\g\bar\b)}=\g^m_{\bar\a\bar\b} A_{\g m}. \label{DAb}
\end{align}

Using the covariant derivatives algebra and gamma matrix identities
these equations imply a chain of equations that define the
supergravity fields as higher components of
$A_{\b\bar\b}(X,\t,\bar\t)$ and put them on-shell. The first few are
given by
\begin{align}
D_\a A_{m\bar\g} -\p_m A_{\a\bar\g} =(\g_m)_{\a\b} W^\b{}_{\bar\g}, \label{WW}\\
D_\a W^\b{}_{\bar\g}=\frac14(\g^{mn})_\a{}^\b F_{mn\bar\g} \label{FF},\\
D_{\bar\a} A_{\g m} -\p_m A_{\g\bar\a} =(\g_m)_{\bar\a\bar\b} W_\g{}^{\bar\b} \label{WWb},\\
D_{\bar\a} W_\g{}^{\bar\b}=\frac14(\g^{mn})_{\bar\a}{}^{\bar\b} F_{\g mn} \label{FFb},
\end{align}
where $F_{mn\bar\g}=\p_{[m} A_{n]\bar\g}$ and $F_{\g mn}=\p_{[m} A_{\g n]}$.

The integrated vertex operator can be seen as a deformation of the
flat space action
\begin{align}
  S_{\rm def} = S_0 +\mu \int\! d^2z\, V,
\label{expansion}
\end{align}
This deformation induces a change of order $\mu$ in the flat space BRST
transformations (\ref{Qleft}) and (\ref{Qright}). We will call the
generator of these new transformations $Q_1$. Invariance of
$S_{\rm def}$ under BRST transformations up to order $\mu$ implies that
\begin{align}\label{defBRST}
  (Q+Q_1)\left(S_0+\mu\int\! d^2z V\right)=Q_1S_0+\mu\int\! d^2z QV=0.
\end{align}
The integrated vertex operator is not uniquely defined; we can make
a non-linear field redefinition of the fundamental fields at order
$\mu$ that will change $V$ by terms proportional to the world-sheet
equations of motion. Since $Q_1S_0$ is proportional to the equations
of motion, we can first solve (\ref{defBRST}) on-shell and require
that
\begin{align}\label{QV1}
  QV=\partial \bar W -\bar\partial W
\end{align}
up to terms proportional to the flat space equations of motion. The
off-shell solution of (\ref{defBRST}) can then be constructed.

Since (\ref{Qleft}) and (\ref{Qright}) are on-shell nilpotent on
all matter fields and combinations of the ghost variables that
are invariant under (\ref{gaugeOmegas}), (\ref{QV1}) implies that
\begin{align}\label{Qw}
  QW=\partial U,\quad Q\bar W =\bar\partial U,
\end{align}
and finally we have that $QU=0$, where $U$ is the unintegrated vertex
operator defined before.

The idea is to find $V$ from $U$ up to world-sheet equations of
motion. We first start with (\ref{Qw}) to find $W$ and $\bar W$.
Consider the equation for $W$ first. Its form can be guessed
knowing that we can use $\partial\bar\l^{\bar\a}=0$ and that it should
have classical dimension $(1,0)$. $W$ turns out to be
\begin{align}\label{W}
  W=\bar\l^{\bar\b} \left( \p\t^\a A_{\a\bar\b}+\Pi^mA_{m\bar\b}+
  d_\a W^\a{}_{\bar\b} +\frac12 N^{mn} F_{mn\bar\b}\right) .
\end{align}
After using the BRST
transformations (\ref{Qleft}) and the equations
\begin{align}
  \bar\l^{\bar\a}\bar\l^{\bar\b}D_{\bar\a} A_{m\bar\b}=
  \bar\l^{\bar\a}\bar\l^{\bar\b}D_{\bar\a}W^\g{}_{\bar\b}=
  \bar\l^{\bar\b}N^{mn}(\l^aD_\a+\bar\l^{\bar\a}D_{\bar\a})F_{mn\bar\b}=0
\end{align}
which are consequences of (\ref{DA}),~(\ref{WW}) and (\ref{FF}), we
obtain that $W$
satisfies $QW=\p U$ . We now prove that (\ref{W}) transforms
in the right way under the residual gauge symmetry from $\d U=Q\L$.
The superfield $A_{\a\bar\b}$ transforms as
\begin{align}
\d A_{\a\bar\b}=D_\a\bar\L_{\bar\b}+D_{\bar\b}\L_\a,
\label{dAab}
\end{align}
where the gauge parameters $\L$ and $\bar\L$ satisfy
(\ref{condL})--(\ref{chainLbar}).
The gauge transformations for the fields in (\ref{W}) come
from their definition in (\ref{DA})--(\ref{FF}). They are
\begin{align}
\d A_{m\bar\b}=\p_m\bar\L_{\bar\b}-D_{\bar\b}\L_m,\quad \d W^\a{}_{\bar\b}=D_{\bar\b}\L^\a,\quad \d F_{mn\bar\b}=-D_{\bar\b}\p_{[m}\L_{n]} .
\label{dchain0}
\end{align}
Using this, $W$ transforms as
\begin{align}
  \d W=\p(\l^\a\L_\a+\bar\l^{\bar\a}\bar\L_{\bar\a})-
  Q\left(\p\t^\a\L_\a+\Pi^m\L_m+d_\a\L^\a+\frac12 N^{mn}\p_{[m}\L_{n]}\right) ,
\label{dW}
\end{align}
as required by $QW=\p U$. Note that the second term closely resembles
the the integrated vertex operator for SYM multiplet in the open
superstring.

Similarly, $\bar W$ can be found using,
\begin{align}
  \l^\a\l^\b D_\a A_{\b m}=\l^\a\l^\b D_\a W_\b{}^{\bar\g}=
  \l^\b \bar N^{mn}(\l^\a D_\a+\bar\l^{\bar\a})F_{\b mn}=0
\end{align}
which are consequences of (\ref{DA})--(\ref{FFb}). We obtain that
\begin{align}\label{Wbar}
\bar W=\l^\a (\bar\p\bar\t^{\bar\b} A_{\a\bar\b}+\bar\Pi^mA_{\a m}+\bar d_{\bar\b} W_\a{}^{\bar\b} +\frac12 \bar N^{mn} F_{\a mn} ) ,
\end{align}
satisfies $Q\bar W=\bar\p U$. As $W$ above, $\bar W$ transforms
adequately under the residual gauge transformation (\ref{dAab}).

The last step is to find $V$ such that $QV=\p\bar W-\bar\p W$.
Knowing that $V$ has to have classical dimension $(1,1)$, vanishing
ghost number and we can ignore term proportional to the world-sheet
equations of motion, we can guess the following form
\begin{align}
V=&~ \p\t^\a \bar\p\bar\t^{\bar\b} A_{\a\bar\b}+\p\t^\a\bar\Pi^mA_{\a m}-\Pi^m\bar\p\bar\t^{\bar\a}A_{m\bar\a}+d_\a\bar\p\bar\t^{\bar\b}W^\a{}_{\bar\b}-\bar d_{\bar\a}\p\t^\b W_\b{}^{\bar\a} \label{ivo0}
\cr &+\frac12\p\t^\a\bar N^{mn} F_{\a
      mn}-\frac12\bar\p\bar\t^{\bar\a}N^{mn}F_{mn\bar\a}+\Pi^m\bar\Pi^n
      A_{mn}+d_\a\bar d_{\bar\b}P^{\a\bar\b}\cr &+\frac14 N^{mn}\bar N^{pq}
      S_{mnpq}+\Pi^m\bar d_{\bar\a}E_m{}^{\bar\a}+d_\a\bar\Pi^m
      E_m{}^\a +\frac12\Pi^m\bar N^{np} \O_{mnp}\cr  &+\frac12
      N^{np}\bar\Pi^m\bar\O_{mnp}+\frac12 d_\a\bar N^{mn}
      C_{mn}{}^\a+\frac12 N^{mn} \bar d_{\bar\a}C_{mn}{}^{\bar\a}.
\end{align}
Using the BRST transformations and $QV= \partial\bar W -\bar\partial
W$ is possible to show that $A_{mn}$ is defined by
\begin{align}
D_{(\a} A_{\b)m}=\g^n_{\a\b}A_{nm},\quad D_{(\bar\a}A_{m\bar\b)}=-\g^n_{\bar\a\bar\b}A_{mn} ,
\label{Amn0}
\end{align}
$E_m{}^{\bar\g}$ and $E_m{}^\g$ are defined by
\begin{align}
D_{(\a}W_{\b)}{}^{\bar\g}=-\g^m_{\a\b}E_m{}^{\bar\g},\quad D_{(\bar\a}W^\g{}_{\bar\b)}=\g^m_{\bar\a\bar\b}E_m{}^\g ,
\label{E0}
\end{align}
$\O$ and $\bar\O$ are defined by
\begin{align}
D_{(\a}F_{\b)mn}=\g^p_{\a\b}\O_{pmn},\quad D_{(\bar\a}F_{mn\bar\b)}=-\g^p_{\bar\a\bar\b}\bar\O_{pmn} ,
\label{O0}
\end{align}
$P$ is defined by
\begin{align}
D_\a E_m{}^{\bar\b}+\p_m W_\a{}^{\bar\b}=-(\g_m)_{\a\g}P^{\g\bar\b},\quad D_{\bar\a} E_m{}^\b-\p_m W^\b{}_{\bar\a}=(\g_m)_{\bar\a\bar\g}P^{\b\bar\g} ,
\label{P0}
\end{align}
$C_{mn}{}^\b$ and $C_{mn}{}^{\bar\b}$ are defined by
\begin{align}
D_\a\O_{pmn}-\p_p F_{\a mn}=(\g_p)_{\a\b}C_{mn}{}^\b,\quad D_{\bar\a}\bar\O_{pmn}+\p_p F_{mn\bar\a}=(\g_p)_{\bar\a\bar\b} C_{mn}{}^{\bar\b} ,
\label{C0}
\end{align}
and finally $S$ is defined by
\begin{align}
D_\a C_{mn}{}^\b=\frac14(\g^{pq})_\a{}^\b S_{pqmn},\quad D_{\bar\a} C_{mn}{}^{\bar\b}=\frac14(\g^{pq})_{\bar\a}{}^{\bar\b} S_{mnpq} .
\label{S0}
\end{align}
It can be shown that these definitions are equivalent among
themselves after writing all the superfields in terms of $A_{\a\bar\b}$
and superspace derivatives. Note that the gauge transformation of $V$
is given by
\begin{align}
  \delta V =& \bar\partial\left(\p\t^\a\L_\a+\Pi^m\L_m+d_\a\L^\a+
              \frac12 N^{mn}\p_{[m}\L_{n]}\right)\cr
              &-\partial\left(\bar\p\bar\t^{\bar\a}\bar\L_{\bar\a}+
              \bar\Pi^m\bar\L_m+\bar d_{\bar\a}\bar\L^{\bar\a}+
              \frac12 \bar N^{mn}\p_{[m}\bar\L_{n]}\right),
\end{align}
and vanishes after integration. There are no BRST exact terms above
because the only way to have well defined covariant operators with
ghost number $-1$ invariant under (\ref{gaugeOmegas}) is using
the non-minimal pure spinor formalism \cite{Berkovits:2005bt}.  In
particular, we can see from (\ref{Amn0}) and (\ref{E0}) that the gauge
transformations of $A_{mn}$, $E_m{}^\a$ and $E_m{}^{\bar\a}$ are given by
\begin{align}
  \delta A_{mn}=\partial_n\L_m - \partial_m\bar\L_n,\quad
  \delta E_m{}^\a = \partial_m \L^\a,\quad
  \delta E_m{}^{\bar\a} = -\partial_m \bar\L^{\bar\a},
\end{align}
which are the expected gauge transformations for the graviton plus the
$B$-field and local supersymmetries for the gravitini. Note that the
$\gamma$-trace of $E_m{}^\a$ and $E_m{}^{\bar\a}$ are invariant by a
consequence of (\ref{chainL}) and (\ref{chainLbar}) and are identified
as the dilatini.

\section{Unintegrated vertex operator in type IIB superstring in $AdS_5\times S^5$}\label{3}

We will now focus on the Type IIB superstring in an $AdS_5\times S^5$
background using the pure spinor description. It is well known that
this background is described by the coset $PSU(2,2|4)/Sp(1,1)\times
Sp(2)$. We will denote the generators of the $\psu(2,2|4)$ algebra by
$T_A=(M_{ab}, M_{a'b'}, P_{\ua}, Q_\a, Q_{\bar\a})$. Their non-zero
(anti-)commutators are
\begin{align}\label{algebra}
[ M_{\uab},M_{\ucd} ] =& -\eta_{\ua[\uc} M_{\ud]\ub} + \eta_{\ub[\uc}
  M_{\ud]\ua},\quad [ P_{\ua} , M_{\ubc} ] =-\eta_{\ua[\ub}
  P_{\uc]},\cr [ P_a , P_b ] =& M_{ab},\quad [ P_{a'} , P_{b'} ] =
  -M_{a'b'},\cr  [M_{\uab},Q_\a]=&\frac12(\g_{\uab})_\a{}^\b Q_\b,\quad
  [M_{\uab},Q_{\bar\a}]=\frac12(\g_{\uab})_{\bar\a}{}^{\bar\b}
  Q_{\bar\b}, \cr [P_{\ua},Q_\a]=&\frac12(\g_{\ua}\eta)_\a{}^{\bar\b}Q_{\bar\b},\quad
  [P_{\ua},Q_{\bar\a}]=-\frac12(\g_{\ua}\eta)_{\bar\a}{}^\b Q_\b,\quad
\{Q_\a,Q_\b\}=\g^{\ua}_{\a\b} P_{\ua},\cr
  \{Q_{\bar\a},Q_{\bar\b}\}=& \g^{\ua}_{\bar\a\bar\b}P_{\ua},\quad
  \{Q_\a,Q_{\bar\b}\}=-\frac12\left((\g^{ab}\eta)_{\a\bar\b}M_{ab}-(\g^{a'b'}\eta)_{\a\bar\b}M_{a'b'}\right).
\end{align}
The non-vanishing structure constants $f_{BC}{}^A$ can read off from
these (anti-)commutators. The geometric quantities are defined
using the Maurer-Cartan currents. They are defined in terms of the coset element $g$ as
\begin{align}\label{geometry}
  (g^{-1}\partial_Mg)^AT_A=J_M{}^A T_A=E_M{}^{\ua}P_{\ua} + E_M{}^\a Q_\a +
  E_M{}^{\bar\a} Q_{\bar\a} +\frac12 \Omega_M{}^{ab}M_{ab}+\frac12\Omega_M{}^{a'b'}M_{a'b'},
\end{align}
where $\partial_M$ are derivatives with respect to local coordinates
$Z^M$.

Additionally, we need the
background values of the RR field-strength and the NSNS two-form,
which are not defined by the geometry,
\begin{align}
P^{\a\bar\b}=-\frac{1}{2} \eta^{\a\bar\b},\quad B_{\a\bar\b}= \eta_{\a\bar\b} ,
\label{PB}
\end{align}
where $\eta_{\a\bar\b}=\g^{01234}_{\a\bar\b},
\eta^{\a\bar\b}=\g_{01234}^{\a\bar\b}$.
We can also calculate the non-zero values of the torsion and the curvature
\begin{align}
T_{\a\b}{}^{\ua}=-\g_{\a\b}^{\ua},\quad T_{\bar\a\bar\b}{}^{\ua}=-\g^{\ua}_{\bar\a\bar\b},\quad T_{\ua\a}{}^{\bar\b}=-\frac{1}{2} (\g_{\ua}\eta)_\a{}^{\bar\b},\quad T_{\ua\bar\a}{}^\b=\frac{1}{2}(\g_{\ua}\eta)_{\bar\a}{}^\b
\label{backTR}\cr
  R_{\a\bar\b\, ab}=(\g_{ab}\eta)_{\a\bar\b},\quad R_{\a\bar\b\, a'b'}=
  -(\g_{a'b'}\eta)_{\a\bar\b},\quad R_{ab\, cd}=-
  \eta_{a[c}\eta_{d]b},\quad R_{a'b'\, c'd'}= \eta_{a'[c'}\eta_{d']b'} .
\end{align}
Here $\ua=(a, a')$ with $a (a')$ refers to the tangent vectorial index
of $AdS_5 (S^5)$. With all these ingredients, the world-sheet action
in the pure spinor formalism for this background is
\begin{align}
S=&\int d^2z\Big[ \frac12 J^{\ua}\bar J^{\ub}
  \eta_{\ua\ub}+\frac{1}{2}  ( J^\a \bar J^{\bar\b} - 3 J^{\bar\b}\bar
  J^\a ) \eta_{\a\bar\b} \cr  &+\o_\a\bar\nabla\l^\a + \bar\o_{\bar\a}
  \nabla \bar\l^{\bar\a} - \frac{1}{2} ( N^{ab} \bar
  N_{ab}-N^{a'b'}\bar N_{a'b'}) \Big] ,
\label{Sads}
\end{align}
where we are using
\begin{align}\label{localCur}
J^A=\partial Z^MJ_M{}^A,\quad \bar J^A=\bar\partial
Z^M J_M{}^A,\quad  N^{\uab}=\frac12(\l\g^{\uab}\o),\quad \bar
  N^{\uab}=\frac12(\bar\l\g^{\uab}\bar\o).
\end{align}
It is important to note that
the action does not depend on all components of ghost currents,
however, fluctuations of these background can and will depend on them.
We have integrated out the $d_\a$ and $\bar d_{\bar\a}$ variables of the
pure spinor formalism. After this, the BRST charge (\ref{Q0}) becomes
\begin{align}
  Q=\oint \left( -2 \l^\a \eta_{\a\bar\b} J^{\bar\b}+
  2\bar\l^{\bar\a}\eta_{\b\bar\a}\bar J^\b \right) .
\label{Qbrst}
\end{align}

The BRST transformation of the coset element
is $Q g=g(\l+\bar\l)$, where we used the shorthand $\l=\l^\a Q_\a$
and $\bar\l=\bar\l^{\bar\a} Q_{\bar\a}$. The Maurer-Cartan
currents $J=g^{-1}\p g$  transform as
\begin{align}
Q J^{ab}&=-(\l\g^{ab}\eta)_{\bar\a} J^{\bar\a} + (\bar\l\g^{ab}\eta)_\a J^\a,\quad Q J^{a'b'}=(\l\g^{a'b'}\eta)_{\bar\a} J^{\bar\a} - (\bar\l\g^{a'b'}\eta)_\a J^\a, \label{QJab}\\
Q J^{\ua}&=(\l\g^{\ua})_\a J^\a+(\bar\l\g^{\ua})_{\bar\a} J^{\bar\a}, \label{QJ}\cr Q J^\a &= \nabla\l^\a -\frac12 (\bar\l\g_{\ua}\eta)^\a J^{\ua},\quad Q J^{\bar\a}=\nabla\bar\l^{\bar\a}+\frac12(\l\g_{\ua}\eta)^{\bar\a} J^{\ua}.
\end{align}
and similarly for $\bar J=g^{-1}\bar\p g$. The pure spinor
ghosts $\l,\bar\l$ are BRST invariant and their conjugate momenta
transform as
\begin{align}
Q\o_\a=-2\eta_{\a\bar\a} J^{\bar\a},\quad Q\bar\o_{\bar\a}=2\eta_{\a\bar\a} \bar J^\a .
\label{Qom}
\end{align}

The structure of the unintegrated vertex operator will be the same as
in the flat space case
\begin{align}
  U=\l^\a \bar\l^{\bar\a} A_{\a\bar\a}(g),
\label{U}
\end{align}
where the superfield $A_{\a\bar\a}(g)$ is a function of the coset
element $g$. The equations coming from $QU=0$ are \cite{Berkovits:2000yr}
\begin{align}
\nabla_{(\a} A_{\b)\bar\g} = \g^{\ua}_{\a\b} A_{\ua\bar\g},\quad \nabla_{(\bar\a} A_{\g\bar\b)}=\g^{\ua}_{\bar\a\bar\b} A_{\g\ua} ,
\label{QU}
\end{align}
where $\nabla_A=
E_A{}^M(\partial_M +\frac12 \Omega_M{}^{ab}M_{ab}+\frac12\Omega_M{}^{a'b'}M_{a'b'})$
for $A=\{\ua,\a,\bar\a\}$ are the covariant derivatives. The
residual gauge symmetry from $\d U=Q\L$ implies that
\begin{align}
\d A_{\a\bar\b}=\nabla_\a\bar\L_{\bar\b}+\nabla_{\bar\b}\L_\a ,
\label{dAads}
\end{align}
where the gauge parameters $\L$ and $\bar\L$ satisfy
\begin{align}
\nabla_{(\a}\L_{\b)}=\g^{\ua}_{\a\b}\L_{\ua},\quad \nabla_{(\bar\a}\bar\L_{\bar\b)}=\g^{\ua}_{\bar\a\bar\b}\bar\L_{\ua} ,
\label{nlab}
\end{align}
which are the generalization of (\ref{condL}) and (\ref{dAab}) for
the $AdS_5\times S^5$ case. As in the flat space case, there are
consequences of these constraint equations. They are
\begin{align}
\nabla_\a\L_{\ua}-\nabla_{\ua}\L_\a=(\g_{\ua})_{\a\b}\L^\b,\quad \nabla_\a\L^\b-\frac12\eta^{\b\bar\g}\nabla_{\bar\g}\L_\a=\frac14(\g^{\uab})_\a{}^\b\nabla_{[\ua}\L_{\ub]} ,
\label{naL}\\
\nabla_{\bar\a}\bar\L_{\ua}-\nabla_{\ua}\bar\L_{\bar\b}=(\g_{\ua})_{\bar\a\bar\b}\bar\L^{\bar\b},\quad \nabla_{\bar\a}\bar\L^{\bar\b}+\frac12\eta^{\g\bar\b}\nabla_\g \bar\L_{\bar\a} =\frac14(\g^{\uab})_{\bar\a}{}^{\bar\b}\nabla_{[\ua}\bar\L_{\ub]} .
\label{Lbar}
\end{align}

 These equations are similar to the equation satisfied by the super-Maxell fields in ten-dimensions, that is (\ref{DA}) and (\ref{DAb}).  Our goal is to find the remaining fields with equations similar to (\ref{WW}), (\ref{FF}) and (\ref{WWb}), (\ref{FFb}). Consider the first equation in (\ref{QU}) . We note that the combination $(\nabla_\a A_{\ua\bar\g}-\nabla_{\ua}A_{\a\bar\g})$ satisfies
\begin{align}
\g^{\ua}_{(\a\b} (\nabla_{\g)} A_{\ua\bar\g}-\nabla_{\ua}A_{\g)\bar\g}) =0 .
\label{restriction}
\end{align}
The proof is the same as in the flat space case \cite{Witten:1985nt}, using the
general  (anti-)commutator of covariant derivatives acting on a
superfield $A_{CD}$
\begin{align}
[\nabla_A,\nabla_B] A_{C D} = -T_{AB}{}^E \nabla_E A_{C D} -R_{AB\,
  C}{}^E A_{E D} - R_{AB\, D}{}^{ E} A_{C E} ,
\label{comm}
\end{align}
the relevant commutators are the same as in flat space. Therefore we
have that
\begin{align}
\nabla_\a A_{\ua\bar\g}-\nabla_{\ua}A_{\a\bar\g}= (\g_{\ua})_{\a\b} W^\b{}_{\bar\g} ,
\label{DAW}
\end{align}
because the identity $(\g^{\ua})_{(\a\b}(\g_{\ua})_{\g)\d}=0$.

The next equation in the chain is for $\nabla_\a W^\b{}_{\bar\g}$.
From (\ref{DAW}), we obtain after some covariant derivative algebra
and the background values for the curvature
\begin{align}
10\nabla_\a W^\b{}_{\bar\g} =\; &(\g^{\uab})_\a{}^\b
F_{\uab\bar\g}-(\g^{\ua})_{\a\r} (\g_{\ua})^{\b\g} \nabla_\g
W^\r{}_{\bar\g} \cr &+ (\g^{\ua})^{\b\g}  \left(
  [\nabla_{\ua},\nabla_\g]A_{\a\bar\g} + [\nabla_{\ua},\nabla_\a]
  A_{\g\bar\g} \right),
\end{align}
where
\begin{align}\label{defF}
  F_{\uba\bar\g}=\nabla_{\ub} A_{\ua\bar\g}-\nabla_{\ua} A_{\ub\bar\g}.
\end{align}
This is again very similar to the flat ten-dimensional superspace \cite{Witten:1985nt},  the difference is in the terms involving commutators. Using (\ref{comm}),
\begin{align}
[\nabla_{\ua},\nabla_\g]A_{\a\bar\g}=-T_{\ua\g}{}^A \nabla_A A_{\a\bar\g}=\frac12(\g_{\ua}\eta)_\g{}^{\bar\d}\nabla_{\bar\d} A_{\a\bar\g} ,\\
[\nabla_{\ua},\nabla_\a]A_{\g\bar\g}=-T_{\ua\a}{}^A \nabla_A
  A_{\g\bar\g}=\frac12(\g_{\ua}\eta)_\a{}^{\bar\d}\nabla_{\bar\d}
  A_{\g\bar\g} ,
\end{align}
we obtain,
\begin{align}
\left( \nabla_\a W^\b{}_{\bar\g} - \frac12 \eta^{\b\bar\d}\nabla_{\bar\d} A_{\a\bar\g} \right) + \frac{1}{10} \g^{\ua}_{\a\r} \g_{\ua}^{\b\g} \left( \nabla_\g W^\r{}_{\bar\g} -\frac12 \eta^{\r\bar\d} \nabla_{\bar\d} A_{\g\bar\g} \right) = \frac{1}{10} (\g^{\uab})_\a{}^\b F_{\uab\bar\g} .
\label{DWis}
\end{align}
This is solved by
\begin{align}
\nabla_\a W^\b{}_{\bar\g} - \frac12 \eta^{\b\bar\d}\nabla_{\bar\d} A_{\a\bar\g}=\frac14(\g^{\uab})_\a{}^\b F_{\uab\bar\g} .
\label{DWF}
\end{align}
Similarly from the second equation in (\ref{QU}), we obtain
\begin{align}
\nabla_{\bar\a}A_{\g\ua}-\nabla_{\ua} A_{\g\bar\a}{}=(\g_{\ua})_{\bar\a\bar\b} W_\g{}^{\bar\b},\quad \nabla_{\bar\a} W_\g{}^{\bar\b} + \frac12 \eta^{\d\bar\b} \nabla_\d A_{\g\bar\a}=\frac14 (\g^{\uab})_{\bar\a}{}^{\bar\b} F_{\g\uab} ,
\label{oth}
\end{align}
where $F_{\g\uab} =\nabla_{\ua} A_{\g\ub}-\nabla_{\ub} A_{\g\ua}$.

Let us summarize our results up to now. The unintegrated vertex
operator is $U=\l^\a\bar\l^{\bar\b}A_{\a\bar\b}$ and its BRST
invariance imply a chain of superfields defined by
\begin{align}
\nabla_{(\a} A_{\b)\bar\g}=\g^{\ua}_{\a\b} A_{\ua\bar\g},\label{DAc}\\
\nabla_\a A_{\ua\bar\g} -\nabla_{\ua} A_{\a\bar\g} =(\g_{\ua})_{\a\b} W^\b{}_{\bar\g}, \label{WWc}\\
\nabla_\a W^\b{}_{\bar\g}-\frac12\eta^{\b\bar\d}\nabla_{\bar\d} A_{\a\bar\g}=\frac14(\g^{\uab})_\a{}^\b F_{\uab\bar\g} \label{FFc},\\
\nabla_{(\bar\a} A_{\g\bar\b)}=\g^{\ua}_{\bar\a\bar\b} A_{\g\ua}, \label{DAbc}\\
\nabla_{\bar\a} A_{\g\ua} -\nabla_{\ua} A_{\g\bar\a} =(\g_{\ua})_{\bar\a\bar\b} W_\g{}^{\bar\b} \label{WWbc},\\
\nabla_{\bar\a} W_\g{}^{\bar\b}+\frac12\eta^{\d\bar\b}\nabla_\d A_{\g\bar\a}=\frac14(\g^{\uab})_{\bar\a}{}^{\bar\b} F_{\g\uab} \label{FFbc},
\end{align}
These equations are the analogous to the equations
(\ref{DA}-\ref{FFb})
of flat superspace.

Before ending this section, we obtain the transformations of the superfields in (\ref{DAc})--(\ref{FFbc}) under the gauge transformations of  (\ref{dAads}). Using (\ref{naL}), (\ref{Lbar}) and commutation relations in our background we obtain
\begin{align}
  \d  A_{\ua\bar\b}&=\nabla_{\ua}\bar\L_{\bar\b}-\nabla_{\bar\b}\L_{\ua}+\frac12(\g_{\ua}\eta)_{\bar\b}{}^\g\L_\g,\cr
  \d W^\g{}_{\bar\b}&=\frac12 \eta^{\gamma\bar\delta}\nabla_{\bar\b}
   \left(\bar\L_{\bar\d} + 2\eta_{\alpha\bar\delta}\L^\a\right)+\frac12(\eta\g^{\ua})^\g{}_{\bar\b}(\L_{\ua}-\bar\L_{\ua}) ,\cr
  \d F_{\uab\bar\b } &= -\nabla_{\bar\b}\nabla_{[\ua}\L_{\ub]} +
     \frac12 (\gamma_{[\ua}\eta\gamma_{\ub]})_{\bar\b\a}\L^\a
     -{1\over 4}R_{\uab\,\ucd}(\gamma^{\ucd})_{\bar\b}{}^{\bar\a}\bar\L_{\bar\a},\label{gauge1}\\
 \d A_{\a\ua}&=
 \nabla_{\ua}\L_\a-\nabla_\a\bar\L_{\ua}-\frac12(\g_{\ua}\eta)_\a{}^{\bar\b}\bar\L_{\bar\b},\cr
 \d W_\a{}^{\bar\g}& =-\frac12\eta^{\d\bar\g}
 \nabla_\a\left(\L_\d-2\eta_{\d\bar\a}\bar\L^{\bar\a}\right)+\frac12(\g^{\ua}\eta)_\a{}^{\bar\g}(\L_{\ua}-\bar\L_{\ua}),\cr
   \d F_{\a\uab } &= -\nabla_{\a}\nabla_{[\ua}\bar\L_{\ub]} -
       \frac12 (\gamma_{[\ua}\eta\gamma_{\ub]})_{\a\bar\b}\bar\L^{\bar\b}
       -{1\over 4}R_{\uab\,\ucd}(\gamma^{\ucd})_{\a}{}^{\b}\L_{\b}.\label{gauge2}
\end{align}
The gauge transformations for $F_{\uab\bar\b}$ and $F_{\a\uab}$
require a separate note. The combination
$(\gamma_{[\ua}\eta\gamma_{\ub]})$ vanishes when one indices is in
$AdS_5$ and the other in $S^5$. Similarly, $R_{\uab\,\ucd}$ is zero
unless all its four indices are in $AdS_5$ or all four are in
$S^5$. In these cases, the transformations simplify to
\begin{align}
  \d F_{ab'\bar\b }=-\nabla_{\bar\b}\nabla_{[a}\L_{b']},\quad
  \d F_{\a ab' }=-\nabla_{\a}\nabla_{[a}\bar\L_{b']}.
\end{align}
On the other hand, when both indices are in, for example,  $AdS_5$,
the transformation for $F_{ab\bar\b}$ is
\begin{align}
  \delta F_{ab\bar\b} = -\nabla_{\bar\b}\nabla_{[a}\L_{b]}
  +\frac12(\gamma_{ab})_{\bar\b}{}^{\bar\a}( \bar\L_{\bar\a} +2\eta_{\a\bar\a}\L^\a).
\end{align}
Combinations of the types
$(\L_\d-2\eta_{\d\bar\a}\bar\L^{\bar\a})$
and $( \bar\L_{\bar\a} +2\eta_{\a\bar\a}\L^\a)$ will
appear again when we discuss the local supersymmetry transformation
for the gravitino, they are the local supersymmetry parameters for the
$AdS_5\times S^5$ superspace.

We will use the transformations the to verify that the analogous of (\ref{W}) and (\ref{Wbar}) in $AdS_5\times S^5$ transform adequately.

\section{Integrated vertex operator in type IIB
  superstring in $AdS_5\times S^5$}\label{4}

After the initial discussion on the superstring in $AdS_5\times S^5$
background, the unintegrated vertex operator and some of its
consequences, we want to explicitly solve the equations
(\ref{QV1}),~(\ref{Qw}) that define the integrated vertex operator
(\ref{expansion}).  Unlike the flat space case, the $b$ ghosts of the
$AdS_5\times S^5$ superstring can be constructed without the
introduction of non-minimal fields \cite{Berkovits:2008ga,Berkovits:2010zz}.
Therefore it is guaranteed that $\p U$ is cohomologically trivial.

As before we will work on-shell. The first equation of motion we will
need is the one for $\bar\l$  obtained from (\ref{Sads})
\begin{align}
\nabla\bar\l^{\bar\a}=\frac14(\bar\l\g^{[ab]})^{\bar\a}N_{[ab]} ,
\label{nlb}
\end{align}
where $A^{[ab]} B_{[ab]}\equiv A^{ab} B_{ab}-A^{a'b'} B_{a'b'}$.

From
the same argument as before, working on-shell allows us to chose $W$
to have the following form
\begin{align}
W=\bar\l^{\bar\b} \left( J^\a A_{\a\bar\b}+J^{\ua} A_{\ua\bar\b}-2\eta_{\a\bar\a}J^{\bar\a}W^\a{}_{\bar\b}+\frac12 N^{\uab} F_{\uab\bar\b} \right) .
\label{Wads}
\end{align}

Acting with $Q$ and using the equations (\ref{QU}), (\ref{DAW}) and (\ref{DWF}) we obtain,
\begin{align}\label{QWads}
QW=&\p U - \l^\a \nabla \bar\l^{\bar\b} A_{\a\bar\b} + \bar\l^{\bar\b}
  (\bar\l\g^{\ua})_{\bar\g} J^{\bar\g} A_{\ua\bar\b} +
  \bar\l^{\bar\b}  \bar\l^{\bar\g} J^{\ua} \nabla_{\bar\g}
  A_{\ua\bar\b} - \frac12\bar\l^{\bar\b} (\bar\l\g_{\ua}\eta)^\a
  J^{\ua} A_{\a\bar\b} \cr
  &-2\eta_{\a\bar\a}  \bar\l^{\bar\b} \nabla  \bar\l^{\bar\a}
  W^\a{}_{\bar\b}+2\eta_{\a\bar\a} \bar\l^{\bar\b} J^{\bar\a}
  \bar\l^{\bar\g} \nabla_{\bar\g} W^\a{}_{\bar\b}\cr
  &+ \frac12 \bar\l^{\bar\b} \l^\g N^{\uab} \nabla_\g F_{\uab\bar\b}
  + \frac12 \bar\l^{\bar\b} \bar\l^{\bar\g} N^{\uab} \nabla_{\bar\g} F_{\uab\bar\b} .
\end{align}
Let us consider first the fourth term above. Using (\ref{dAads}),
(\ref{comm}), (\ref{backTR}) and
\begin{align}
  \gamma_{ab}\gamma_{\underline{cdefg}}\gamma^{ab}-
  \gamma_{a'b'}\gamma_{\underline{cdefg}}\gamma^{a'b'}=0,\quad
  \g^{[bc]}\g_a\g_{[bc]}=16\g_a,\quad \g^{[bc]}\g_{a'}\g_{[bc]}=-16\g_{a'},
\end{align}
we obtain
\begin{align}
\bar\l^{\bar\b} \bar\l^{\bar\g} \nabla_{\bar\b} A_{\ua\bar\g}=\frac12 \bar\l^{\bar\b} (\bar\l\g_{\ua}\eta)^\a A_{\a\bar\b}.
\label{llA}
\end{align}
which cancels with the last two terms in the first line of
(\ref{QWads}). Consider now the second term in the second
line of (\ref{QWads}), it contains
\begin{align}
\bar\l^{\bar\b} \bar\l^{\bar\g} \nabla_{\bar\b} W^\a{}_{\bar\g} &= \frac{1}{10} \bar\l^{\bar\b} \bar\l^{\bar\g} (\g^{\ua})^{\a\d} \nabla_{\bar\b} \left( \nabla_\d A_{\ua\bar\g} - \nabla_{\ua} A_{\d\bar\g} \right)\cr
&=\frac{1}{10} \bar\l^{\bar\b} \bar\l^{\bar\g} (\g^{\ua})^{\a\d}
  \left( \{ \nabla_\d , \nabla_{\bar\b} \} A_{\ua\bar\g} - \nabla_\d
  \nabla_{\bar\b}A_{\ua\bar\g} + [ \nabla_{\ua} , \nabla_{\bar\b} ]
  A_{\d\bar\g} - \nabla_{\ua}\nabla_{\bar\b} A_{\d\bar\g} \right) .
\end{align}
Using (\ref{comm}) and (\ref{llA}),
\begin{align}
  \bar\l^{\bar\b} \bar\l^{\bar\g} \nabla_{\bar\b} W^\a{}_{\bar\g} =
  \frac{1}{10}\bar\l^{\bar\b} \bar\l^{\bar\g} (\g^{\ua})^{\a\d} \left(
  - R_{\d\bar\b\ua}{}^{\ub} A_{\ub\bar\g} -\frac12
  (\g_{\ua}\eta)_{\bar\b}{}^\r \nabla_\d A_{\r\bar\g} -
  T_{\ua\bar\b}{}^\r \nabla_\r A_{\d\bar\g} \right).
\end{align}
Finally, after using (\ref{backTR}) we obtain
\begin{align}
\bar\l^{\bar\b} \bar\l^{\bar\g} \nabla_{\bar\b} W^\a{}_{\bar\g} = -\frac12 \bar\l^{\bar\b} (\bar\l\g^{\ua}\eta)^\a A_{\ua\bar\g} ,
\label{llW}
\end{align}
which helps to cancel the second term in the second line with the third term in the first line of (\ref{QWads}).

Up to now we have
\begin{align}
QW=\p U - \l^\a \nabla \bar\l^{\bar\b} A_{\a\bar\b} - 2\eta_{\a\bar\a}  \bar\l^{\bar\b} \nabla  \bar\l^{\bar\a} W^\a{}_{\bar\b} + \frac12 \bar\l^{\bar\b} \l^\g N^{\uab} \nabla_\g F_{\uab\bar\b} + \frac12 \bar\l^{\bar\b} \bar\l^{\bar\g} N^{\uab} \nabla_{\bar\g} F_{\uab\bar\b} .
\label{QWadsUp}
\end{align}
The last term can be written as
\begin{align}
\bar\l^{\bar\b} \bar\l^{\bar\g} \nabla_{\bar\b} F_{\uab\bar\g} =\frac12 \bar\l^{\bar\b} \bar\l^{\bar\g} (\g_{[\ua}\eta\g_{\ub]})_{\bar\b\r}W^\r{}_{\bar\g},
\label{llF}
\end{align}
after using the definition of $F_{\uab\bar\gamma}$ (\ref{defF}) and
commuting derivatives.
Note that $\g_{[\ua}\eta\g_{\ub]}$ is different from zero only if $(\ua,\ub)=(a,b)$ or $(\ua,\ub)=(a',b')$, then  $\frac12 N^{\uab} \g_{[\ua}\eta\g_{\ub]} = N^{[ab]} \g_{[ab]}\eta$. The equation (\ref{QWadsUp}) reduces to
\begin{align}
QW=&\p U - \l^\a \nabla \bar\l^{\bar\b} A_{\a\bar\b}  + \frac12 \bar\l^{\bar\b} \l^\g N^{\uab} \nabla_\g F_{\uab\bar\b}\cr &- 2\eta_{\a\bar\a}  \bar\l^{\bar\b} \left( \nabla \bar\l^{\bar\a} -\frac14\bar\l^{\bar\g} N^{[ab]}(\g_{[ab]})_{\bar\g}{}^{\bar\a}   \right)W^\a{}_{\bar\b}.
\label{QWadsUpP}
\end{align}
Note that the last term vanishes on-shell. It remains to compute $\nabla_\g F_{\uab\bar\b}$. Using (\ref{DAW}) and commuting derivatives,
\begin{align}
\nabla_\g F_{\uab\bar\b} = -(\g_{[\ua})_{\g\r} \nabla_{\ub]} W^\r{}_{\bar\b} - \frac12(\g_{[\ua}\eta)_\g{}^{\bar\d} \nabla_{\bar\d} A_{\ub]\bar\b} - R_{\uab\g}{}^\d A_{\d\bar\b} - R_{\uab\bar\b}{}^{\bar\d} A_{\g\bar\d} .
\label{DF}
\end{align}
From here and using $\l^\g N^{\uab}(\g_{\ua})_{\g\d}=-\frac12 (\l\o) (\l\g^{\ub})_\d$ we obtain
\begin{align}
  \l^\g N^{\uab} \nabla_\g F_{\uab\bar\b} =& (\l\o)(\g^{\ua})_{\g\r}
  \left( \nabla_{\ua} W^\r{}_{\bar\b} + \frac12 \eta^{\r\bar\d}
  \nabla_{\bar\d} A_{\ua\bar\b} \right)\cr  &- \l^\g N^{\uab} \left(
  R_{\uab\g}{}^\d A_{\d\bar\b} + R_{\uab\bar\b}{}^{\bar\d}
  A_{\g\bar\d} \right) .
\end{align}
The first term in this expression contains
\begin{align}(\g^{\ua})_{\a\b} \nabla_{\ua} W^\b{}_{\bar\g} = \{
  \nabla_\a , \nabla_\b \} W^\b{}_{\bar\g},
\end{align}
after using  (\ref{DWF}) and (\ref{DF}) we can show
\begin{align}
(\g^{\ua})_{\a\b} \left( \nabla_{\ua} W^\b{}_{\bar\g} + \frac12 \eta^{\b\bar\r} \nabla_{\bar\r} A_{\ua\bar\g} \right) = 0 ,
\label{coupledRReq}
\end{align}
therefore, (\ref{QWadsUpP}) simplifies to
\begin{align}\label{QWcasi}
QW =& \p U - 2\eta_{\a\bar\a}  \bar\l^{\bar\b} \left( \nabla \bar\l^{\bar\a} -\frac14\bar\l^{\bar\g} N^{[ab]}(\g_{[ab]})_{\bar\g}{}^{\bar\a}   \right)W^\a{}_{\bar\b}
- \l^\a \nabla \bar\l^{\bar\b} A_{\a\bar\b}\cr
&-\frac12 \bar\l^{\bar\b} \l^\g N^{\uab} \left( R_{\uab\g}{}^\d
      A_{\d\bar\b} + R_{\uab\bar\b}{}^{\bar\d} A_{\g\bar\d} \right).
\end{align}
After using the background values of the curvature (\ref{backTR})
\begin{align}
N^{\uab} R_{\uab\a}{}^\b =  -\frac12 N^{[ab]} (\g_{[ab]})_\a{}^\b,\quad N^{\uab} R_{\uab\bar\a}{}^{\bar\b} =  -\frac12 N^{[ab]} (\g_{[ab]})_{\bar\a}{}^{\bar\b} ,
\end{align}
we finally obtain
\begin{align}
QW = \p U - \left( \nabla \bar\l^{\bar\a} -\frac14\bar\l^{\bar\g} N^{[ab]}(\g_{[ab]})_{\bar\g}{}^{\bar\a}   \right) \left( 2\eta_{\a\bar\a}  \bar\l^{\bar\b}  W^\a{}_{\bar\b} + \l^\a A_{\a\bar\a}  \right) .
\label{QWF}
\end{align}
Using the equation of motion (\ref{nlb}) we have shown that
\begin{align}
QW=\p U .
\label{QWispU}
\end{align}
A similar calculation shows that the operator $\bar W$ that satisfies
$Q\bar W = \bar\p U$ is given by
\begin{align}
\bar W=\l^\a \left( \bar J^{\bar\b} A_{\a\bar\b} + \bar J^{\ua} A_{\a\ua} + 2\eta_{\b\bar\b} \bar J^\b W_\a{}^{\bar\b} + \frac12 \bar N^{\uab} F_{\a\uab} \right) .
\label{WbarADS}
\end{align}
Performing the same steps one can show that
\begin{align}
Q\bar W =  \bar\p U - \left( \bar\nabla\l^\a -\frac14 \l^\g \bar N^{[ab]} (\g_{[ab]})_\g{}^\a \right) \left( 2\eta_{\a\bar\b} \l^\g W_\g{}^{\bar\b} + \bar\l^{\bar\b} A_{\a\bar\b} \right) .
\label{QbarW}
\end{align}
Note that the second term is proportional to the equation of motion for $\l$.

In summary, we have found the conformal weights $(1,0)$ and $(0,1)$ superfields
\begin{align}
W=\bar\l^{\bar\b}\left( J^\a A_{\a\bar\b}+J^{\ua}A_{\ua\bar\b}-2\eta_{\a\bar\a} J^{\bar\a} W^\a{}_{\bar\b}+\frac12 N^{\uab} F_{\uab\bar\b}\right),\label{WadsS} \\\bar W=\l^\a\left( \bar J^{\bar\b} A_{\a\bar\b}+\bar J^{\ua} A_{\a\ua}+2\eta_{\b\bar\b}\bar J^\b W_\a{}^{\bar\b}+\frac12\bar N^{\uab} F_{\a\uab}\right) \label{WbaradsS} ,
\end{align}
which satisfy $QW=\p U$ and $Q\bar W=\bar\p U$ on-shell.

From the definition of $W$, is has to transform under (\ref{dAads}) as
\begin{align}
  \delta W = \partial(\l^\a\L_\a+\bar\l^{\bar\a}\bar\L_{\bar\a}) - Q(\Psi),
\end{align}
as in flat space. Using the gauge transformations of the superfields
inside $W$, we obtain that
\begin{align}
\d W=\partial(\l^\a\L_\a+\bar\l^{\bar\a}\bar\L_{\bar\a})-Q\left(J^\a\L_\a+J^{\ua}\L_{\ua}-2\eta_{\a\bar\a}J^{\bar\a}\L^\a+\frac12 N^{\uab}\nabla_{[\ua}\L_{\ub]}\right) .
\label{dWads}
\end{align}
which fixes $\Psi$ to be
\begin{align}\label{Psi}
  \Psi= J^\a\L_\a+J^{\ua}\L_{\ua}-2\eta_{\a\bar\a}J^{\bar\a}\L^\a+
  \frac12 N^{\uab}\nabla_{[\ua}\L_{\ub]}.
\end{align}
An analogous result holds for $\bar W$
\begin{align}
\d \bar W=\bar\partial(\l^\a\L_\a+\bar\l^{\bar\a}\bar\L_{\bar\a})-Q(\bar\Psi),
\label{dWbarads}
\end{align}
with $\bar\Psi$ given by
\begin{align}\label{Psibar}
  \bar\Psi= \bar J^{\bar\a}\bar\L_{\bar\a}+\bar J^{\ua}\bar \L_{\ua}
  +2\eta_{\a\bar\a}\bar J^{\a}\bar\L^{\bar\a}+
  \frac12 \bar N^{\uab}\nabla_{[\ua}\bar\L_{\ub]}.
\end{align}

It is interesting to note
that this looks like the vertex operator for  $N=1$ vector multiplet in ten
dimensions for an open superstring, but now in an $AdS_5\times S^5$
background. Perhaps this identification can be made more precise
by studying boundary conditions for the pure spinor string in this
space. Appropriate boundary conditions will reduce the number of
independent components of $J^\a$ and $J^{\bar\a}$.

Now that $W$ and $\bar W$ have been constructed, we can find the
conformal weight $(1,1)$ operator $V$ that satisfies
$QV=\p\bar W-\bar\p W$.
We will need to use the equations of motion of the world-sheet
action and some Maurer-Cartan identities. Also, the equations for the
pure spinor ghosts imply that the ghost Lorenz currents satisfy
\begin{align}
\bar\nabla N^{ab}=- N^{c[a} \bar N_c{}^{b]},\quad \bar\nabla N^{a'b'}= N^{c'[a'} \bar N_{c'}{}^{b']},\quad \bar\nabla N^{ab'}= N^{c'a}\bar N_{c'}{}^{b'}+ \bar N^{ca} N_c{}^{b'} \label{dbarN},\\
\nabla\bar N^{ab} = N^{c[a} \bar N_c{}^{b]},\quad \nabla\bar N^{a'b'}=- N^{c'[a'} \bar N_{c'}{}^{b']},\quad \nabla\bar N^{ab'}= N^{ca}\bar N_c{}^{b'}+ \bar N^{c'a} N_{c'}{}^{b'}. \label{dNbar}
\end{align}
The equations of morion for the currents $J^A$ for $A=(\ua,\a,\bar\a)$ are
\begin{align}
\bar\nabla J^{\bar\a} =& \frac14 N^{[ab]} \bar J^{\bar\b}
   (\g_{[ab]})_{\bar\b}{}^{\bar\a}+\frac14\bar N^{[ab]} J^{\bar\b}
   (\g_{[ab]})_{\bar\b}{}^{\bar\a},\cr
   \nabla\bar J^\a=&\frac14 N^{[ab]} \bar J^\b
   (\g_{[ab]})_\b{}^\a+\frac14\bar N^{[ab]} J^\b
   (\g_{[ab]})_\b{}^\a, \label{dJferm} \cr
   \bar\nabla J^{\ua}=& -\g^{\ua}_{\bar\a\bar\b} J^{\bar\a}\bar J^{\bar\b} -
                        \d^{\ua}_b ( N^{bc}\bar J_c +\bar N^{bc} J_c ) +
                        \d^{\ua}_{b'} (N^{b'c'}\bar J_{c'}+\bar N^{b'c'} J_{c'} ),\cr
   \nabla\bar J^{\ua}=& \g^{\ua}_{\a\b} J^\a \bar J^\b - \d^{\ua}_b (
   N^{bc}\bar J_c +\bar N^{bc} J_c ) + \d^{\ua}_{b'}
   (N^{b'c'}\bar J_{c'}+\bar N^{b'c'} J_{c'} ).
\end{align}
They are calculated varying the action with respect to $\delta g=
g(\delta X^{\ua}T_{\ua}+\delta X^\a T_\a+ \delta X^{\bar\a}T_{\bar\a})$
and using the Maurer-Cartan identities
\begin{align}
  \partial \bar J^A -\bar\partial J^A - J^B \bar J^Cf_{CB}{}^A =0.
\end{align}

The vertex operator $V$ couples all gauge covariant currents with
conformal weight $(1,0)$ with the ones with weight $(0,1)$ and it has
the form
\begin{align} \label{Vads}
V=\; &2\eta_{\b\bar\g}J^\a\bar J^\b
  W_\a{}^{\bar\g}-2\eta_{\g\bar\a}J^{\bar\a}\bar
  J^{\bar\b}W^\g{}_{\bar\b} + J^\a \bar J^{\bar\b} A_{\a\bar\b} + J^\a
  \bar J^{\ua} A_{\a\ua} -J^{\ua}\bar J^{\bar\a} A_{\ua\bar\a} \cr
  &+\frac12 J^\a \bar N^{\uab} F_{\a\uab}-\frac12 N^{\uab}\bar
    J^{\bar\a} F_{\uab\bar\a}+J^{\bar\b} \bar J^\a\bar {\cal
    V}_{\a\bar\b}+J^{\ua}\bar J^\a \bar{\cal V}_{\ua\a}+J^{\bar\a}\bar
    J^{\ua} {\cal V}_{\ua\bar\a}+J^{\ua}\bar J^{\ub} {\cal
    V}_{\uab}\cr
  &+\frac14 N^{\uab}\bar N^{\ucd}{\cal V}_{\uab~\ucd}+
    \frac12 N^{\uab}\bar J^\a \bar{\cal V}_{\uab\a}+
    \frac12 J^{\bar\a} \bar N^{\uab} {\cal V}_{\uab\bar\a}+
    \frac12 J^{\ua}\bar N^{\ubc} {\cal V}_{\ua~\ubc}+
    \frac12 N^{\ubc} \bar J^{\ua}\bar {\cal V}_{\ua~\ubc}.
\end{align}

From the structure of $W$, $\bar W$ and the BRST transformations of
the currents $J^\a$ and $\bar J^{\bar\a}$ we can immediately identify the
first seven superfields in $V$ with the ones that already appeared in $W$ and
$\bar W$. The remaining superfields, collectively called ${\cal V}$,
are constrained by $QV=\partial\bar W-\bar\partial W$.

For example, $\bar{\cal V}_{\ua\a}$ is constrained  by the equation
\begin{align}
\g^{\ua}_{\b\g}\bar{\cal V}_{\ua\a}=\g^{\ua}_{\a(\b} A_{\g)\ua} -2\eta_{\a\bar\g}\nabla_{(\b} W_{\g)}{}^{\bar\g} .
\label{V0}
\end{align}
Using the definition (\ref{WWbc}) for $W_\g{}^{\bar\g}$
and (anti-)commuting covariant derivatives, one can obtain
that $\eta_{\a\bar\g}\nabla_{(\b} W_{\g)}{}^{\bar\g}=\frac12
\g^{\ua}_{\a(\b} A_{\g)\ua} + \g^{\ua}_{\b\g} {\cal O}_{\ua\a}$ where
${\cal O}_{\ua\a}$ is a combination of superfields and derivatives.
This allows us to find $\bar{\cal V}_{\ua\a}$. The resulting
expression contains the superfield $A_{\a\bar\b}$ and its
derivatives. The full expression for $\bar{\cal V}_{\ua\a}$ is not
very illuminating and it turns out to be
\begin{align}\label{Vaa}
\bar{\cal V}_{\ua\a}=\; &-\frac{1}{10}  (\eta\g^{\ub}\g_{\ua}\eta)_\a{}^\b
    A_{\b\ub}+\frac{1}{40}(\eta\g_{\ua})_\a{}^{\bar\b}
     \eta^{\g\bar\d}\nabla_{\bar\b} A_{\g\bar\d} \cr
  & -\frac{1}{10}(\eta\g^{\ub})_\a{}^{\bar\b}
    \left( \nabla_{\ub}
    A_{\ua\bar\b}+\frac18(\eta\g_{\ua}\g_{\ub})^{\g\bar\d}
    \nabla_{\bar\b} A_{\g\bar\d}-\frac14\g_{\ua}^{\g\d}\nabla_{\bar\b}A_{\g\bar\d}\right).
\end{align}

All other superfields $\cal V$ in (\ref{Vads}) will appear in
equations that can be inverted. We will write all their defining
equations. Their explicit expressions can be obtained in the same way
as the one for $\bar{\cal V}_{\ua\a}$ in  (\ref{Vaa}).

The superfields ${\cal V}_{\ua\bar\a}$ satisfies the equation
\begin{align}
\g^{\ua}_{\bar\b\bar\g}{\cal V}_{\ua\bar\a}=-\g^{\ua}_{\bar\a(\bar\b} A_{\ua\bar\g)} -2\eta_{\g\bar\a}\nabla_{(\bar\b} W^{\g}{}_{\bar\g)} .
\label{V1}
\end{align}

The superfield ${\cal V}_{\uab}$ which, from its coupling in
(\ref{Vads}), can be identified with fluctuations of the metric plus
the NSNS two-form, satisfies the equations
\begin{align}
\g^{\ub}_{\a\b}{\cal V}_{\uba}=\nabla_{(\a} A_{\b)\ua}+\frac12(\g_{\ua}\eta)_{(\a}{}^{\bar\g} A_{\b)\bar\g},\quad \g^{\ub}_{\bar\a\bar\b} {\cal V}_{\uab}=-\nabla_{(\bar\a} A_{\ua\bar\b)}+\frac12(\g_{\ua}\eta)_{(\bar\a}{}^\g A_{\g\bar\b)} ,
\label{V2}
\end{align}
which are equivalent when they are expressed in terms of
$A_{\a\bar\b}$. Note that ${\cal V}_{\uab}$ contains the components
${\cal V}_{ab'}$ that twist the $AdS_5$ and $S^5$ spaces.

The superfield $\bar{\cal V}_{\a\bar\b}$ satisfies the equations
\begin{align}\label{V3}
\frac12(\g_{\ua}\eta)_\a{}^{\bar\g} \bar{\cal V}_{\b\bar\g}+\frac12(\g_{\ua}\eta)_\b{}^{\bar\g} A_{\a\bar\g} +\g^{\ub}_{\a\b}{\cal V}_{\uab} -\nabla_\a\bar{\cal V}_{\ua\b}-2\eta_{\b\bar\g}\nabla_{\ua}W_\a{}^{\bar\g} = 0 ,\cr
\frac12(\g_{\ua}\eta)_{\bar\a}{}^\g \bar{\cal V}_{\g\bar\b}+\frac12(\g_{\ua}\eta)_{\bar\b}{}^\g A_{\g\bar\a} +\g^{\ub}_{\bar\a\bar\b}{\cal V}_{\uba} -\nabla_{\bar\a}{\cal V}_{\ua\bar\b}-2\eta_{\g\bar\b}\nabla_{\ua}W^\g{}_{\bar\a} = 0 ,
\end{align}
which are also equivalent when written in terms of $A_{\a\bar\b}$.

The next four superfields are related to fluctuations of the connections. First, ${\cal V}_{\uab\bar\b}$ is given by
\begin{align}
\frac12(\g^{\uab}\eta)_{\bar\a\g}{\cal V}_{\uab\bar\b} +\frac12(\g^{\uab}\eta)_{\g\bar\b}F_{\uab\bar\a}-\nabla_{\bar\a}\bar{\cal V}_{\g\bar\b} +\nabla_{\bar\b}A_{\g\bar\a} -\g^{\ua}_{\bar\a\bar\b}\bar{\cal V}_{\ua\g}=0,
\label{V4}
\end{align}
and $\bar{\cal V}_{\uab\a}$ is determined by
\begin{align}
\frac12(\g^{\uab}\eta)_{\a\bar\b}\bar{\cal V}_{\uab\g} +\frac12(\eta\g^{\uab})_{\g\bar\b}F_{\a\uab}-\nabla_{\a}\bar{\cal V}_{\g\bar\b} +\nabla_{\g}A_{\a\bar\b} +\g^{\ua}_{\a\g}{\cal V}_{\ua\bar\b}=0,
\label{V5}
\end{align}
and ${\cal V}_{\ua~\ubc}$ satisfies the equation
\begin{align}
\frac12(\g^{\ubc}\eta)_{\bar\a\b}{\cal V}_{\ua~\ubc}+\nabla_\b A_{\ua\bar\a} -\nabla_{\bar\a}\bar{\cal V}_{\ua\b}-(\g_{\ua}\eta)_{\bar\a}{}^\g\eta_{\b\bar\g}W_\g{}^{\bar\g}= 0.
\label{V6}
\end{align}
Finally, the background field $\bar{\cal V}_{\ua~\ubc}$ satisfies the equation
\begin{align}
\frac12(\g^{\ubc}\eta)_{\a\bar\b}\bar{\cal V}_{\ua~\ubc}+\nabla_{\bar\b} A_{\a\ua} +\nabla_\a{\cal V}_{\ua\bar\b}-(\g_{\ua}\eta)_\a{}^{\bar\g}\eta_{\g\bar\b}W^\g{}_{\bar\g}= 0.
\label{V7}
\end{align}
The last superfield to be defined is related to fluctuations of the curvature ${\cal V}_{\uab~\ucd}$ and it satisfies the equations
\begin{align}\nonumber
\bar N^{\uab}\left( \frac12(\g^{\ucd}\eta)_{\a\bar\b}{\cal V}_{\ucd~\uab}+\nabla_{\bar\b}F_{\a\uab} + \nabla_\a {\cal V}_{\uab\bar\b} \right)\cr +\frac12\bar N^{[ab]}\left( (\g_{[ab]})_\a{}^\g\bar{\cal V}_{\g\bar\b}+(\g_{[ab]})_{\bar\b}{}^{\bar\g} A_{\a\bar\g}\right) &= 0 ,
\end{align}
\begin{align}
N^{\uab}\left( \frac12(\g^{\ucd}\eta)_{\bar\a\b}{\cal V}_{\uab~\ucd}+\nabla_\b F_{\uab\bar\a} - \nabla_{\bar\a} \bar{\cal V}_{\uab\b} \right)\cr +\frac12 N^{[ab]}\left( (\g_{[ab]})_{\bar\a}{}^{\bar\g}\bar{\cal V}_{\b\bar\g}+(\g_{[ab]})_\b{}^\g A_{\g\bar\a}\right) &= 0 .
\label{V8}
\end{align}

As a simple application one can calculate the integrated vertex operator (\ref{Vads}) for the radius operator found in \cite{Berkovits:2008ga}. The unintegrated vertex operator is $U=\l^\a \bar\l^{\bar\b}\eta_{\a\bar\b}$ and the integrated vertex operator should be proportional to the lagrangian in (\ref{Sads}). Using $A_{\a\bar\a}=\eta_{\a\bar\a}$ the
solution for (\ref{V0})-(\ref{V8}) gives
\begin{align}
V=J^\a\bar J^{\bar\b} \eta_{\a\bar\b} - 3 J^{\bar\b}\bar J^\a \eta_{\a\bar\b} + 2J^{\ua}\bar J^{\ub} \eta_{\uab} + N^{ab}\bar N_{ab} - N^{a'b'}\bar N_{a'b'} .
\label{Vr}
\end{align}
Which is proportional to the lagrangian up to the equations of motion for the ghosts.

\subsection{Local symmetries and field content}

Due to the extra mixing coming from the RR background, non vanishing torsions and curvatures, it is quite involved to get the equations of motion for the superfields defined by $V$ in a way that we can identify what is their lowest component. However, we can use the gauge invariance and look for fields that transform in the expected way.
As in flat space, $V$ inherits a gauge transformation from the
unintegrated vertex operator  $\delta U =
Q(\l^\a\L_\a+\bar\l^{\bar\a}\bar\L_{\bar\a})$, after finding the gauge
transformations of $W$ and $\bar W$, we have that $V$ transforms as
\begin{align}\label{deltaVtotal}
  \delta V =\, & \bar\partial
  \left(J^\a\L_\a+J^{\ua}\L_{\ua}-2\eta_{\a\bar\a}J^{\bar\a}\L^\a+\frac12
  N^{\uab}\nabla_{[\ua}\L_{\ub]}\right) \cr
  &-\partial\left(\bar J^{\bar\a}\bar\L_{\bar\a}+\bar
  J^{\ua}\bar\L_{\ua}+2\eta_{\a\bar\a}\bar J^{\a}\bar\L^{\bar\a}+\frac12
  \bar N^{\uab}\nabla_{[\ua}\bar\L_{\ub]}\right),
\end{align}
the integral of vertex operator (\ref{Vads}) is gauge invariant under (\ref{dAads}).

The transformation of each field in $V$ (\ref{Vads}) can be found
using the definition above and the equations for the Maurer-Cartan
currents and $N^{\uab}$, $\bar N^{\uab}$.  Alternatively, we can use
the definition of each $\cal V$ in terms of the superfields in $W$ and
$\bar W$ and use (\ref{dAads}), (\ref{gauge1}) and (\ref{gauge2}).

We can find a very simple expression for the superfield whose lowest
component gives the metric plus the NSNS two-form
\begin{align}
  {\cal V}_{\uab}= {1\over 16}(\eta\g_{\ua}\g_{\ub})^{\a\bar\a}A_{\a\bar\a}
  -{1\over 64} \g_{\ua}^{\bar\a\bar\b}\g_{\ub}^{\a\b}
  \nabla_{\bar\a}\nabla_\a A_{\b\bar\b}.
\end{align}
Its gauge transformation can be computed and it is given by
\begin{align}
  \delta {\cal V}_{\uab}= \nabla_{b}\Lambda_{a} - \nabla_{a}\bar\Lambda_{b}.
\end{align}
where we can see that $\L_a -\bar\L_a$ is related to the
diffeomorphism parameter and $\L_a + \bar\L_a$ is the parameter of the
gauge transformation for the NSNS two-form.

The superfield whose lowest component is the gravitino can be
identified by looking for the correct gauge transformation. It turns
out that the combination $\bar{\cal V}_{\ua\a}-
\frac14 (\gamma_{\ua} {}^{\ub} )_\a{}^\b A_{\b \ub}$ transforms as
\begin{align}
  \delta \left( \bar{\cal V}_{\ua\a}-\frac14 (\gamma_{\ua} {}^{\ub} )_\a{}^\b A_{\b \ub}\right) =
  \frac12 (\gamma_{\ua}\eta)_\a{}^{\bar\b}(\bar\Lambda_{\bar\b} +
  2\eta_{\delta\bar\b} \Lambda^\delta) +\nabla_{\ua}(\Lambda_\a
  -2\eta_{\a\bar\gamma} \bar\Lambda^{\bar\gamma}).
\end{align}
From this we will identify one of the gravitini as the lowest
component of the superfield
\begin{align}
  {\cal E}_{\ua}{}^{\bar\alpha} = \eta^{\a\bar\alpha}\left( \bar{\cal V}_{\ua\a}-\frac14 (\gamma_{\ua}{}^{\ub})_\a{}^\b A_{\b \ub}\right),
\end{align}
which will transform as
\begin{align}
  \delta {\cal E}_{\ua}{}^{\bar\alpha} =
  \frac12 (\eta\gamma_{\ua}\eta)^{\bar\a\bar\b}(\bar\Lambda_{\bar\b} +
  2\eta_{\delta\bar\b} \Lambda^\delta) +\nabla_{\ua}(\eta^{\a\bar\alpha}\Lambda_\a
  -2\bar\Lambda^{\bar\a}).
\end{align}
Note that $(\eta\gamma_{\ua}\eta)$ is $\gamma_a$ and $-\gamma_{a'}$,
which is the expected behavior for the term depending on the
cosmological constant in the local transformation of the
gravitino. Furthermore, it reduces to the usual local supersymmetry
transformation in the limit where the RR background goes to zero.

Now we want to find the bi-spinor superfield whose lowest component is
bi-spinor RR field-strength. Comparing with the flat superstring we
start with ${\cal V}_{\a\bar\a}$ and find that it transforms as
\begin{align}
  \delta {\cal V}_{\a\bar\a} =
  2\eta_{\a\bar\b}\nabla_{\bar\a}\bar\L^{\bar\b} -
  2\eta_{\b\bar\alpha}\nabla_\a \L^\b.
\end{align}
After using (\ref{naL}) and (\ref{Lbar}) we have that\footnote{We
  would like to thank Nathan Berkovits for pointing out an error in
  original version of the derivation below. We also thank him for
  discussions on the correct gauge invariant superfield.}
\begin{align}
\delta ( {\cal V}_{\a\bar\a} + A_{\a\bar\a} ) =  -\frac12 (\g^{\uab}\eta)_{\a\bar\a} \N_{[\ua}\L_{\ub]} - \frac12 (\eta\g^{\uab})_{\a\bar\a} \N_{[\ua} \Lb_{\ub]}.
\label{new1}
\end{align}
Consider ${\cal V}_{\uab~\ucd}$ in (\ref{Vads}). Under gauge transformations it changes as
\begin{align}
\d {\cal V}_{\uab~\ucd}=-2\left( \N_{[\ue}\L_{\ua]} R^{\ue}{}_{\ub\uc\ud} - \N_{[\ue} \L_{\ub]} R^{\ue}{}_{\ua\uc\ud} \right) + 2 \left( R_{\ua\ub\uc}{}^{\ue} \N_{[\ud}\Lb_{\ue]} - R_{\ua\ub\ud}{}^{\ue} \N_{[\uc} \Lb_{\ue]} \right) ,
\label{new4}
\end{align}
where $R$ is the space-time curvature given in (\ref{backTR}). This
can be seen from (\ref{deltaVtotal}) and the equations of motion for
the ghost Lorentz currents
(\ref{dbarN}) and (\ref{dNbar}) compactly writen using the curvature. It turns out that the combination $C_{\uab}=\eta^{\ucd}V_{\ua\uc~\ub\ud}$, will transform as
\begin{align}
&\d C_{ab}=-6 ( \N_{[a} \L_{b]} + \N_{[a} \Lb_{b]} )=-6\d {\cal V}_{[ab]},\quad \d C_{a'b'}=6 ( \N_{[a'} \L_{b']} + \N_{[a'} \Lb_{b']} )=6\d{\cal V}_{[a'b']}, \cr
&\d C_{ab'}=8 ( \N_{[a} \L_{b']} - \N_{[a} \Lb_{b']} ),\quad \d C_{a'b}=8 ( \N_{[a'} \L_{b]} + \N_{[a'} \Lb_{b]} ),
\label{new2}
\end{align}
where ${\cal V}_{[\uab]}={\cal V}_{\uab}-{\cal V}_{\uba}$. Using this in (\ref{new1}) we obtain that
\begin{align}
  {\cal F}_{\a\bar\a}=
  {\cal V}_{\a\bar\a} + A_{\a\bar\a} + (\g^{ab}\eta)_{\a\bar\a} {\cal V}_{ab} + (\g^{a'b'}\eta)_{\a\bar\a} {\cal V}_{a'b'} + \frac18 (\g^{aa'}\eta)_{\a\bar\a} C_{aa'} ,
   \label{new3}
\end{align}
is gauge invariant.

The usual definition using upper indices can
be achieved using the $\eta$ tensors.  The zero mode of the gauge
invariant scalar $\phi= \eta^{\a\bar\a} {\cal F}_{\a\bar\a}$ is the
operator that changes the radius of the $AdS_5\times  S^5$ space. The
superfields whose lowest components are the dilatini can be found in
similar ways.

\section{Conclusion and prospects}\label{5}

In this paper we did the explicit construction of the integrated
vertex operator in an $AdS_5\times S^5$ background using the pure
spinor formalism. We have found how all superfields present in the
integrated vertex are related to the initial superfield
$A_{\a\bar\a}(g)$ in the unintegrated vertex operator. The analysis
done here complements the formal construction previously done in
\cite{Chandia:2013kja}. However, the integrated vertex operator found
in that work does not depend on the Lorentz ghost currents with mixed
indices. It would be interesting to find the origin of this
discrepancy.

The final answer resembles the construction of the integrated vertex
operator in a flat background, replacing the flat space
supersymmetric currents $(\Pi^m,\partial \theta^\a, d_\a,N^{mn})$ with
the the isometry invariant currents
$(J^{\ua},J^\a,J^{\bar\a},N^{\uab})$. The superspace description of
linearized supergravity found here should be related to the
supergravity pre-potential found in \cite{Polacek:2016nry}.
A possible future application is
to find explicit expressions for vertex operators for some specific
supergravity state defined by its $\psu(2,2|4)$ labels, in the same
way single trace operators of $N=4$ SYM are defined. Progress in
this problem was made in \cite{Berkovits:2012ps} where  explicits
states were found expanding the vertex operator close to the boundary
of $AdS_5$. These results were later used in \cite{Azevedo:2014rva} to
compute amplitudes with one closed string state and open string
states. Another known explicit vertex operator is the one
corresponding to the beta deformation \cite{Bedoya:2010qz}. A possible
way to describe more general explicit states is to use the formalism
described in \cite{Ramirez:2015rca}.

Another interesting direction is to study vertex operators for
open strings in $AdS_5\times S^5$. As we noticed before, the gauge
parameters for the local symmetries of the supergravity states satisfy
equations that resembles the equations of motion of an $N=1$ vector
multiplet in ten dimensions. Of course this identification has to be
made more precise since it is known that there are no D9-branes in
$AdS_5\times S^5$. The boundary interaction of a D-brane
should be an operator of the form
\begin{align}
  I_{\rm boundary} = \int \left( A_{\ua}(g)\sJ^{\ua} + A_\a(g) \sJ^\a
  + A_{\bar\a}(g) \sJ^{\bar\a} + F_{\uab}(g) \sN^{\uab} \right),
\end{align}
where $(A_{\ua}, A_\a ,A_{\bar\a},F_{\uab})$ describe the
massless states of the D-brane and
$(\sJ^{\ua},\sJ^\a,\sJ^{\bar\a},\sN^{\uab})$ are appropriate isometry
invariant world-sheet one-forms evaluated at the boundary. A
BRST analysis of the boundary conditions with the
interaction $I_{\rm boundary}$ along the lines of
\cite{Berkovits:2002ag} should give the correct physical state
conditions for $(A_{\ua}, A_\a ,A_{\bar\a},F_{\uab})$ and the
allowed D-branes.

Perhaps the most important problem is to understand the quantum
correction to the physical states conditions. For the case of
superstrings in flat space, since it is a free theory, the primary state
condition receives only a one loop quantum correction, the anomalous
dimension for the exponential operator $e^{ik\cdot X}$. From the BRST
cohomology point of view there is no correction at all. This will not
be true for fluctuations around a general background since the
physical state condition will receive corrections from vertices of the
action for both computations. However, for the case of
$AdS_5\times S^5$ the situation for the  massless spectrum should
be similar to the flat space case. These states are dual to
the protected BPS operators in the $N=4$ SYM theory.
The only quantum correction to the primary state condition should be
a one loop correction, proportional to the quadratic
Casimir of $\psu(2,2|4)$ algebra \cite{Ramirez:2015rma}.
The BRST cohomology condition should
receive no correction. It would be very interesting to understand how
all possible quantum corrections to the classical calculations in this
work cancel. We plan to address this in a future work.

\vskip 0.2in
{\bf Acknowledgements}~
The work of B{\sc cv} is partially supported by FONDECYT
grant number 1151409 and  CONICYT grant number DPI20140115.

\appendix

\section{The remaining equations}
In this appendix we list the additional equations the superfields
$\cal V$ satisfy. They are all consequences of (\ref{DA}) when written
in terms of $A_{\a\bar\a}$. Some of the equations are written without
removing the Maurer-Cartan or ghost currents since this is their most
compact form. The equations are
\begin{align}
\nabla_\a{\cal V}_{\uab}-\nabla_{\ua} A_{\a\ub}+\frac12(\g_{\ua}\eta)_\a{}^{\bar\b}{\cal V}_{\ub\bar\b}-\frac12(\g_{\ub}\eta)_\a{}^{\bar\b}A_{\ua\bar\b} = 0 ,
\label{R1}
\end{align}
\begin{align}
\nabla_{\bar\a}{\cal V}_{\uab}+\nabla_{\ub}A_{\ua\bar\a}-\frac12(\g_{\ua}\eta)_{\bar\a}{}^\b A_{\b\ub}-\frac12(\g_{\ub}\eta)_{\bar\a}{}^\b\bar{\cal V}_{\ua\b} = 0 ,
\label{R2}
\end{align}
\begin{align}
\l^\a N^{\uab} ( \nabla_\a \bar {\cal V}_{\uab\b}-\g^{\uc}_{\a\b}\bar{\cal V}_{\uc\uab} ) + \l^\a N^{[ab]}(\g_{[ab]})_{\b\bar\g} W_\a{}^{\bar\g} = 0 ,
\label{R3}
\end{align}
\begin{align}
\bar\l^{\bar\a}\bar N^{\uab} ( \nabla_{\bar\a}{\cal V}_{\uab\bar\b}-\g^{\uc}_{\bar\a\bar\b}{\cal V}_{\uc\uab})-\bar\l^{\bar\a}\bar N^{[ab]}(\g_{[ab]}\eta)_{\bar\b\g} W^\g{}_{\bar\a} = 0 ,
\label{R4}
\end{align}
\begin{align}
N^{\uab}(\nabla_\b F_{\uab\bar\a}+\frac12(\g^{\ucd}\eta)_{\bar\a\b}{\cal V}_{\uab\ucd}-\nabla_{\bar\a}\bar{\cal V}_{\uab\b})+\frac12 N^{[ab]}((\g_{[ab]})_{\bar\a}{}^{\bar\g}\bar{\cal V}_{\b\bar\g}+(\g_{[ab]})_\b{}^\g A_{\g\bar\a} ) = 0 ,
\label{R5}
\end{align}
\begin{align}
\bar N^{\uab}(\nabla_{\bar\b} F_{\a\uab}+\frac12(\g^{\ucd}\eta)_{\a\bar\b}{\cal V}_{\ucd\uab}+\nabla_\a {\cal V}_{\uab\bar\b})+\frac12 \bar N^{[ab]}((\g_{[ab]})_{\a}{}^{\g}\bar{\cal V}_{\g\bar\b}+(\g_{[ab]})_{\bar\b}{}^{\bar\g} A_{\a\bar\g} ) = 0 ,
\label{R6}
\end{align}
\begin{align}
\l^\a N^{\uab} \nabla_\a F_{\uab\bar\b}-\frac12\l^\a N^{[ab]}(\g_{[ab]})_{\bar\b}{}^{\bar\g} A_{\a\bar\g} = 0 ,
\label{R7}
\end{align}
\begin{align}
\bar\l^{\bar\a}\bar N^{\uab} \nabla_{\bar\a} F_{\b\uab}-\frac12\bar\l^{\bar\a} \bar N^{[ab]}(\g_{[ab]})_{\b}{}^{\g} A_{\g\bar\a} = 0 ,
\label{R8}
\end{align}
\begin{align}
N^{\uab} ( \nabla_{(\bar\a} F_{\uab\bar\b)}+\g^{\uc}_{\bar\a\bar\b}\bar{\cal V}_{\uc\uab})+N^{[ab]}(\g_{[ab]}\eta)_{(\bar\a\g}W^\g{}_{\bar\b)} = 0 ,
\label{R9}
\end{align}
\begin{align}
\bar N^{\uab} ( \nabla_{(\a} F_{\b)\uab}-\g^{\uc}_{\a\b}{\cal V}_{\uc\uab})+\bar N^{[ab]}(\g_{[ab]}\eta)_{(\a\bar\g}W_{\b)}{}^{\bar\g} = 0 ,
\label{R10}
\end{align}
\begin{align}
\l^\a N^{\uab} \bar J^{\uc} ( -\frac14(\g_{\uc}\eta)_\a{}^{\bar\b} F_{\uab\bar\b}+\frac12\nabla_\a\bar{\cal V}_{\uc\uab})-\l^\a N^{ab} \bar J_a A_{\a b}+\l^\a N^{a'b'}\bar J_{a'}A_{\a b'} = 0 ,
\label{R11}
\end{align}
\begin{align}
\bar\l^{\bar\a}\bar N^{\uab}  J^{\uc} ( -\frac14(\g_{\uc}\eta)_{\bar\a}{}^{\b} F_{\b\uab}+\frac12\nabla_{\bar\a}{\cal V}_{\uc\uab})+\bar\l^{\bar\a}\bar N^{ab} J_a A_{b\bar\a}-\bar\l^{\bar\a}\bar N^{a'b'} J_{a'}A_{b'\bar\a } = 0 ,
\label{R12}
\end{align}
\begin{align}
N^{\uab}\bar J^{\uc} ( \nabla_{\bar\a}\bar{\cal V}_{\uc\uab}+\nabla_{\uc}F_{\uab\bar\a}-\frac12(\g_{\uc}\eta)_{\bar\a}{}^\b \bar{\cal V}_{\uab\a})\cr+\frac12 N^{[ab]}\bar J^{\uc}(\g_{[ab]})_{\bar\a}{}^{\bar\b}{\cal V}_{\uc\bar\b}+N^{ab}\bar J_{[a} A_{b]\bar\a}-N^{a'b'}\bar J_{[a'} A_{b']\bar\a}& =0 ,
\label{R13}
\end{align}
\begin{align}
\bar N^{\uab} J^{\uc} ( \nabla_{\a}{\cal V}_{\uc\uab}-\nabla_{\uc}F_{\a\uab}+\frac12(\g_{\uc}\eta)_{\a}{}^{\bar\b} {\cal V}_{\uab\bar\b})\cr +\frac12 \bar N^{[ab]} J^{\uc}(\g_{[ab]})_{\a}{}^{\b}\bar{\cal V}_{\uc\a}-\bar N^{ab} J_{[a} A_{\a b]}+\bar N^{a'b'} J_{[a'} A_{\a b']}&=0 ,
\label{R14}
\end{align}
\begin{align}
\l^\a N^{\uab}\bar N^{\ucd} \nabla_\a {\cal V}_{\uab\ucd} +\frac12 \l^\a N^{\uab}\bar N^{[ab]}(\g_{[ab]})_\a{}^\b \bar{\cal V}_{\uab\b}
-\l^\a N^{ca}\bar N_c{}^b F_{\a ab}&\cr + \l^\a N^{c'a'}\bar N_{c'}{}^{b'}F_{\a a'b'}-\l^\a N^{ca}\bar N_c{}^{b'}F_{\a ab'}+\l^\a N_{c'}{}^{b'}\bar N^{c'a}F_{\a ab'} &= 0 ,
\label{R15}
\end{align}
\begin{align}
\bar\l^{\bar\a} N^{\uab}\bar N^{\ucd} \nabla_{\bar\a} {\cal V}_{\uab\ucd} +\frac12 \bar\l^{\bar\a} N^{[ab]}\bar N^{\uab}(\g_{[ab]})_{\bar\a}{}^{\bar\b} {\cal V}_{\uab\bar\b}
-\bar\l^{\bar\a} N^{ca}\bar N_c{}^b F_{ab\bar\a}&\cr + \bar\l^{\bar\a} N^{c'a'}\bar N_{c'}{}^{b'}F_{a'b'\bar\a}+\bar\l^{\bar\a} N^{c'a}\bar N_{c'}{}^{b'}F_{ab'\bar\a}+\bar\l^{\bar\a} N_{c}{}^{b'}\bar N^{ca}F_{\a ab'} &= 0.
\label{R16}
\end{align}

{\small
\bibliographystyle{abe}
\bibliography{mybib}{}
}
\end{document}